# A Novel Solution of Using Mixed Reality in Bowel and Oral and Maxillofacial Surgical Telepresence: 3D Mean Value Cloning algorithm

**Arjina Maharjan**[1]**, Abeer Alsadoon**[1*]**, P.W.C. Prasad**[1]**, Nada AlSallami**[2]**, Tarik A. Rashid**[3]**, Ahmad Alrubaie**[4] **, Sami Haddad**[5,6]

[1]School of Computing and Mathematics, Charles Sturt University, Sydney Campus, Australia
[2]Computer Science Department, Worcester State University, MA, USA
[3]Computer Science and Engineering, University of Kurdistan Hewler, Erbil, KRG, Iraq.
[4]Faculty of Medicine, University of New South Wales, Sydney, Australia
[5]Department of Oral and Maxillofacial Services, Greater Western Sydney Area Health Services, Australia
[6]Department of Oral and Maxillofacial Services, Central Coast Area Health, Australia

**Abeer Alsadoon**[1*]
* Corresponding author. A/Prof  (Dr) Abeer Alsadoon, [1]School of Computing and Mathematics, Charles Sturt University, Sydney Campus, Australia, Email: aalsadoon@studygroup.com , Phone +61 2 9291 9387

## Abstract

**Background and aim**: Most of the Mixed Reality models used in the surgical telepresence are suffering from discrepancies in the boundary area and spatial-temporal inconsistency due to the illumination variation in the video frames. The aim behind this work is to propose a new solution that helps produce the composite video by merging the augmented video of the surgery site and the virtual hand of the remote expertise surgeon. The purpose of the proposed solution is to decrease the processing time and enhance the accuracy of merged video by decreasing the overlay and visualization error and removing occlusion and artefacts. **Methodology**: The proposed system enhanced the mean value cloning algorithm that helps to maintain the spatial-temporal consistency of the final composite video. The enhanced algorithm includes the 3D mean value coordinates and improvised mean value interpolant in the image cloning process, which helps to reduce the sawtooth, smudging and discolouration artefacts around the blending region. **Results: As compared to the state of the art solution, the** accuracy in terms of overlay error of the proposed solution is improved from 1.01mm to 0.80mm whereas the accuracy in terms of visualization error is improved from 98.8% to 99.4%. The processing time is reduced to 0.173 seconds from 0.211 seconds. **Conclusion:** Our solution helps make the object of interest consistent with the light intensity of the target image by adding the space distance that helps maintain the spatial consistency in the final merged video.

## 1. Introduction

Some of the surgery's complications require expertise surgeons with outstanding surgical knowledge and skill. Since these expertise surgeons are comparatively less in number globally, it will be time-consuming for them to be present at the surgery site [1]. Therefore, several technologies such as stereo audio, telementoring, telemedicine system, real-time 3D video vision, etc. are developed to overcome these complications. Though these technologies help in remote collaboration, the major limitations of these technologies are the lack of physicality of expert surgeons in the surgery site. As a result, they are unable to help when they needed  [2].

Telepresence surgery is developed to overcome the limitations of these traditional technologies. It allows the expert surgeons of remote place to assist in the surgery of patient at the local site [3]. Mixed reality is one of the latest technologies used in the field of telesurgery in recent times. The use of mixed reality visualization technique enables the collaboration of the local and remote site expert surgeon for the real-time surgeries by improving the accuracy of the merged videos. Mixed reality visualisation is the technique in which augmented reality and virtual reality are merged to create an environment in which both real and virtual objects can interact [4]. However, the technique is limited only in the research and completely theoretical. The technique has not been used practically in the surgical telepresence of oral and maxillofacial surgery due to many issues such as lack of proper evaluation of this technique in the medical field, lack of consideration of the end-users as well as overlay and visualisation errors that affect the accuracy of the merged video.

This work aims to merge the augmented video produced in the local site with the virtual hand of the expert surgeon to produce the augmented virtuality view. This will improve the visualisation of the surgical site and the accuracy of the merged video. As a result, this will support the remote collaboration between the two surgeons for real-time surgeries. The proposed solution enhanced the mean value cloning that helps maintain the spatial-temporal consistency. The 3D mean value coordinates are used to overcome the issue of over blending and spatial-temporal inconsistencies. Mean value interpolation is enhanced by adding the local smoothing term to maintain spatial consistency. Besides, our proposed solution improves the processing time and accuracy of the final composite video by decreasing the overlay and visualization.

The rest of this paper is divided into 4 sections: section 2 gives a literature review and the state of the art solution, section 3 explain the proposed solution and the area of improvements, section 4 provides the results of the proposed system and discussion, and section 5 outlines the conclusions.

## 2. Literature review

The main purpose of this literature review is to gain in-depth knowledge about surgical telepresence, mixed reality, augmented reality and virtual reality. It provides information about the tools, techniques used in these areas. The most significant works in the area of Tri map extraction and alpha matting, and Video cloning and merging process are analyzed and discussed in this section.

Henry & Lee  [5] implemented the automatic tri-map generation algorithm that consists of lazy snapping, graph cut segmentation and Fuzzy c-mean clustering to generate accurate and reliable trimap automatically. The solution helps to minimize the Sum of Absolute Difference (SAD) value and computing time for the tri-map generation and identification of unknown area. This process can improve the quality of alpha matte and minimizing the artefacts. The proposed solution minimized the SAD value from 11,410 to 10,784 and also reduced the computational time for tri-map generation and alpha matte calculation from 13.96 to 5.49 seconds and 11.99 to 6.07 seconds respectively as compared to its state of art solution [6]. Henry & Lee [5] claimed that their approach does not rely on depth data. However, it will be difficult to produce alpha mattes if the image consists of a complicated background. The deep learning technique can be used to process the image containing a complicated background.

Li, Yuan & Fan [7] uses Fuzzy c-means clustering along with density peaks clustering. Density peaks clustering can produce tri maps automatically without any involvement of the user thereby improving the quality of the alpha matte by minimizing the artefacts. It improves the accuracy of matting. Using this proposed solution, the accuracy of matting is enhanced by reducing the Root Mean Squared Error (RMSE) and Mean Absolute Deviation (MAD) value from 5.5 to 1.52 and from 5.18 to 1.02 respectively as compared to its state of art solution [8]. The process of tri map generation still requires human interactions which will result in more processing time. Automatic generation of tri-map should be considered to generate precise tri maps.

Cai et al.  [9] enhanced the earthworm optimization algorithm by enhancing the reproduction and the Cauchy mutation. As a result, the accuracy of the matting is increased. This system improved the quality of matting by reducing the Mean Squared Error (MSE) from 0.0042 to 0.0031 as compared to its state of art solution [10]. However, the process of tri map generation requires human interactions which will need more processing time. The future work should focus on the fully automatic generation of the tri-map, which will help to generate precise tri maps.

Amin, Riaz and Ghafoor [11] introduced a hybrid defocused region segmentation technique that uses the slope of the magnitude spectrum, total variation and local binary patterns as a measure to produce the sharpness maps. This helps to distinguish between the blurred region and the in-focus region of the image. Tri maps are produced and propagated by segmenting the maps which help to detect the pixels of the blurred images. The system provides the precision and recall of 0.87 and 0.55 respectively that is better than the state of the art system which is 0.9 and 0.05 respectively [12]. However, this research did not consider any methods to improve the selective criterion of the sample set. Aldo this research did not explain how to maintain the balance between the region and the precision.

Chen, He & Yu [13] introduced a full feature coverage sampling method. This method uses the edge of the image to cover the most features of this image. Also, it uses particle swarm optimization to maintain the balance between the population of the sample set and the time cost of searching to improve the accuracy of the matting results. This research improves the SAD and MSE from 24.2 to 15.1 and 24.1 to 12.7 respectively as compared to its state of art solution [14]. However, the research did not consider a method to improve the selective criterion of the sample set. Also, it did not study how to maintain the balance between the region and the precision.

Cho, Tai & Kweon [15] introduced a deep convolutional neural network matting, which incorporated closed formed matting and KNN matting to generate high-quality mattes by identifying the local features of the input image. This work has improved the SAD and MSE from 0.1492 to 0.0353 and from 26.0425 to 11.3167 respectively as compared to its state of art solution [16]. However, it has not used noisy and blurred images to estimate the alpha mattes. Future work should focus on estimating the quality alpha matted for such types of images and removing artefacts from those images.

Fan et al.  [17] introduced a hierarchical image matting model to efficiently and effectively improve the accuracy of the blood vessel segmentation, and also reduce the time consumed. This system improves the blood vessel segmentation accuracy from 95.1 to 96% and decreases the processing time from 15.74 to 10.72 seconds as compared to its state of art solution [18]. However, this system has not used noisy and blurred images to estimate the alpha mattes. Future work should focus on estimating the quality alpha matted for noisy and blurred images.

Lin & Chuang [19] introduced a sampling-based matting method and a fully connected CRF method. In these two methods, the best sample sets are collected using cost and spatial distance, and the alpha mattes which estimated initially are corrected to increase the accuracy of quality of the alpha matte. The system improved the accuracy of the alpha matting by reducing the SAD from 16.8 to 7.6, Gradient Error from 16.9 to 13.9 and MSE from 15.6 to 10.4 as compared to its state of art solution [10]. However, this study has not used noisy and blurred images to estimate the alpha mattes. Future work should focus on removing artefacts and estimating the quality alpha matted for those type of images.

Venkata et al. [3] enhanced the mean value cloning algorithm.  This solution used the mean value coordinates for the image cloning process and interpolated every pixel of the image to enhance the accuracy by decreasing overlay error, increasing the visibility and decreasing the processing time. The system improved the video accuracy by 0.40 mm (in terms of the overlay error) and the visibility of pixels from 98.4% to 99.1% as compared to its state of art solution [20]. The system improved the processing time from 11 seconds per 50 frames to 10 seconds, thereby speeding up the surgery time. However, spatial consistency has not been considered in this solution. When the colour difference between the object of interest and the target image is very high, the composited video will not seem realistic. This happens because the blending object will become inconsistent with the brightness of the target video frames. To solve this issue, spatial consistency should be considered.

Shakya et al. [4] implemented the improvised CAMSHIFT algorithm along with the  RGBD cameras and improvised volumetric-based image synthesisation. This solution improved the depth perception accuracy by minimizing the overlay error and processing time. This could result in recovering the occluded region of the video frames. It provided an accuracy of 1.28 mm (overlay error) and a processing time enhanced to 76 sec as compared to its state of the art [21]. However, the gradient domain technique used for image cloning in this solution requires frame by frame supervision which becomes labour intensive.

 Wang et al. [22] introduced the illumination-aware video composition algorithm in which mean-value cloning, interpolation and blending boundary are optimized and illumination guided gradients are merged to obtain flawless composition results. This solution minimized the smudging, the discoloured artefacts and the processing time even though there is an illumination variation in the source and the target videos. The system reduced the pre-processing time by 21.48 seconds per frames and the blending time by 2.8 seconds per frames as compared to its state of art solution [23].  However, occlusion handling was not considered in this work. However, handling is needed to recover the occluded region.

Pokhrel et al. [24] introduced the volume subtraction technique. This work can identify the shape and the structure for accurate cutting, which helps to minimize the cutting error. The proposed solution implemented Iterative Closest Point (ICP) algorithm. ICP consists of a rotational matrix and translation vector that helps to minimize the geometric error thereby improving the registration accuracy. Pokhrel et al. [24] used optical cameras to capture the real-time videos which help to minimize the processing time. This solution improved the overlay accuracy by 0.40~0.55 mm and the processing time of 12-13 frames per second as compared to its state of art solution [25]. However, the computation consuming of the detection part of the Tracking Learning Detection (TLD) algorithm is comparatively huge.

Basnet et al. [26] implemented an enhanced ICP algorithm which consists of a rotational matrix and translation vector that helps to minimize the geometric error. [26] used an optical camera to capture the real-time videos which do not need any re-calibration and maintenance thereby decreasing the processing time. This system implements TLD for the image matching and ICP algorithm along with the Rotational Matrix and Translation Vector (RMaTV) algorithm for the accurate registration of the surgical tools. It improved the accuracy by 0.23~0.35 mm (overlay error) and gives the processing rate of 8~12 frames per second as compared to its state of art solution [25]. However, the cutting error is not considered in this work. Along with the overlay error, the cutting error should also be considered to solve the issues of navigation with augmented reality when cutting the bones in a scheduled direction and depth.

Murugesan et al. [25] introduced a rotational matrix and a translation vector algorithm. This work aims to enhance the ICP algorithm, minimize the geometric error and improve the registration accuracy. This work used stereo cameras to capture the real-time videos which help to improve the depth perception of the image. The proposed system implemented the TLD technique for image matching. This solution improved the accuracy by 0.30~0.40 mm (overlay error) and the processing time of 10-13 frames per second as compared to its state of art solution which provides an accuracy of 0.40mm and processing time of 13 frames per second  [27]. However, the cutting error was not measured in this work. Cutting error should be considered to solve the issues of navigation with augmented reality when cutting bones in a planned direction and depth.

Hu et al. [28] introduced the illumination-aware live videos background replacement algorithm.  The Colored Locality Sensitive Histogram algorithm was used for efficient background segmentation that is robust to illumination variation. The blocked real-time matting was used to produce the accurate alpha mattes that have a smooth boundary. Also, a local antialiasing method was used to improve the distortions on the edges which helps to make the background replacement more flawless. The system has an F1 score of 0.790875 which is higher than that of the state-of-art solution which is 0.7089 [29]. The processing time in this work is computed based on frame size i.e. 100 frames per second for the frame size of 176*144, 35.7 frames per second for 352*288 and 19.8 frames per second for 640*360. However, the accuracy in terms of the overlay error, visualisation error and the occlusion recovery were not considered in this work. They should be considered to make the video composition more flawless and smoother.

## 2.1 State of the art solution

Venkata el at. [3] proposed an enhanced Multi-Layer Mean Value Cloning (EMLMV) algorithm to improve the accuracy by decreasing the overlay error and visualisation error and minimize the processing time. Tri-map and alpha mattes are produced to eliminate the smudging and discolouration artefacts. Venkata el at. [3] introduced volumetric image syncretization in the enhanced multi-layer mean value cloning algorithm to eliminate the occlusion by recovering the background pixels. It provides an accuracy of 0.40 mm (in terms of overlay error, visibility of pixels of 99.1% and a processing time of 10 seconds per 50 frames. This work has three stages, as shown in Figure 1 i.e. Source/Expert side, Target/Surgery side and Mixed reality. Figure 1 shows the features (highlighted inside the blue dotted rectangle) and limitations of the state of art solution (highlighted inside the red dotted rectangle).

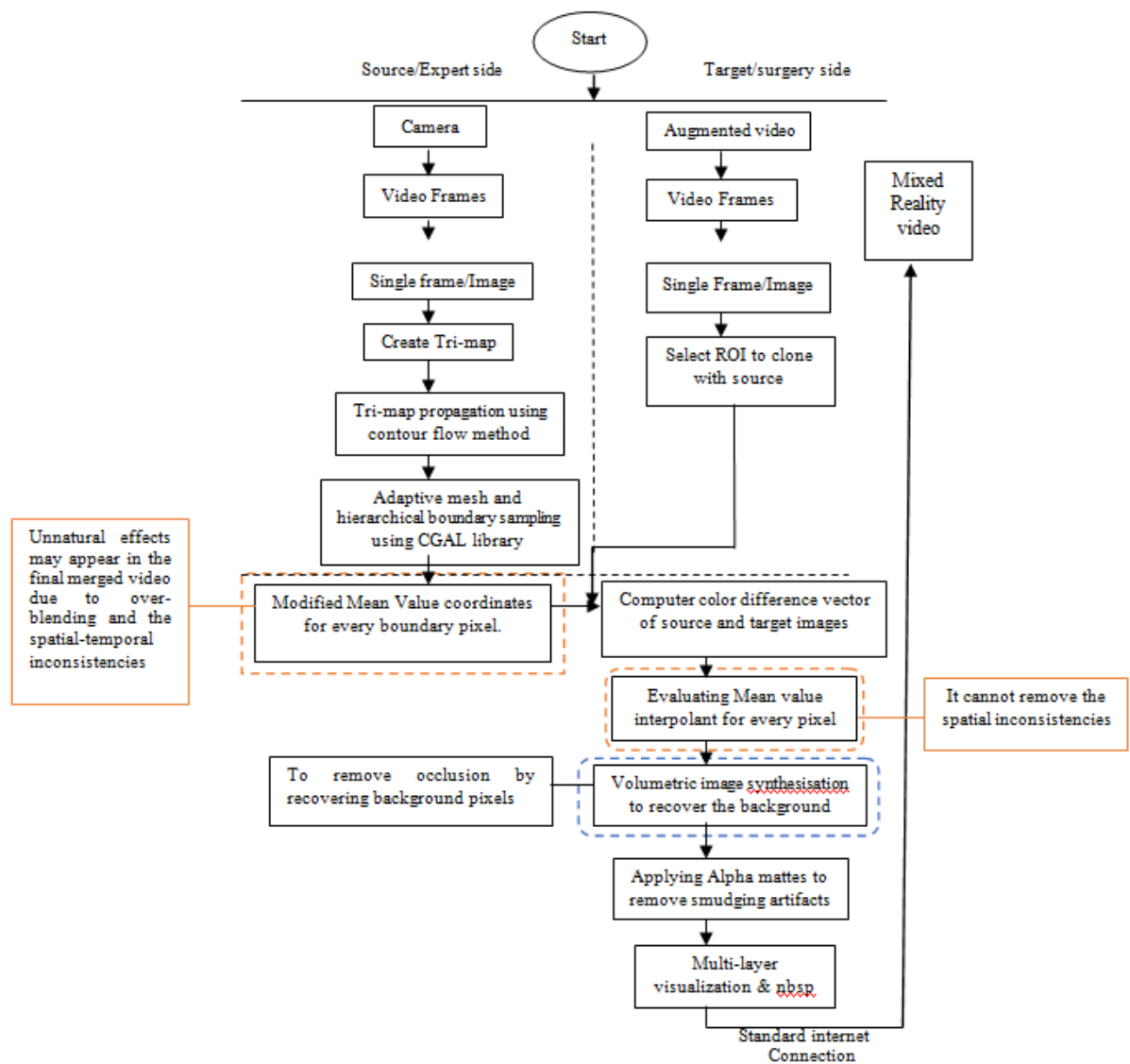

Figure 1: Block Diagram of the State of Art Solution, [3]

(Good features are represented in the rectangles with blue borders and limitations are represented with the rectangles with a red border of the state of art solution)

***Surgery site:*** The Computed Tomography (CT) scan data is used. From the CT scan data, augmented video is produced through segmentation. Video frames are generated from the videos and sent to the expert site through an internet connection to be merged with the expert surgeon's hand video [3].

***Expert surgeon site:*** Real-time surgical video is taken using the camera. Video frames are produced from the recorded video of the guidelines given by the expertise surgeon. Trimaps are generating from the video frames. A tri map propagation technique is used along with the contour flow method to track the hand motion of expert surgeon. A mesh is constructed to the areas on the basis of the Computational Geometry Algorithm Library (CGAL) and hierarchical boundary sampling method is implemented which evaluates and improves the pixels of boundary to process in the later stage [3].

***Mixed reality:*** The mean value coordinates and interpolation coefficient is calculated for every pixel which helps to remove the discolouration artefacts. Then, the mean value interpolation is calculated which shows the discrepancy between the foreground and background pixels. The volumetric image synthesisation method is used to recover the area occluded by the hands of the expert surgeon. Finally, composite results are achieved by combining the synthesized image pixels with the target image pixels, alpha mattes are applied to remove the smudging artefact [3].

**Limitation Justification:** When 2D mean value coordinates are applied in the video composition, unnatural effects may appear in the final merged video due to over-blending and spatial-temporal inconsistencies. Besides, when the colour difference between the object of interest and the target image is very high, the composited video will not seem realistic because the blending object will become inconsistent with the brightness of the target video frames leading to spatial inconsistencies **This means, if** spatial consistency is not considered, it may produce the illumination variation in the final composite video which may increase the visualisation error and thereby decreasing the accuracy. The system proposed by Venkata et al. [3] improved the accuracy of the final composited video by 0.40 mm (in terms of the overlay error) and the visibility of pixels from 98.4% to 99.1%. The system also improved the processing time from 11 seconds per 50 frames to 10 seconds per 50 frames thereby speeding up the surgery time. The enhanced multi-layer mean value cloning is implemented to improve the accuracy in terms of visualization error, overlay error and processing time [3].  However, the final merged video is still affected by the sawtooth artefact and spatial-temporal inconsistencies. **Table 1 shows the** Enhanced multi-layer mean value cloning algorithm proposed by Venkata et al. [3].

The final enhanced multi-layer composite video is calculated using Eq. (1) [3]:

$$I\ layers(p) = \alpha\ I_c(p) + \beta\ I_{mc}(p) \qquad (1)$$

Where,
p = Pixel
$\alpha, \beta$ = Blending parameters
$I_c(p)$ = Cloned region without smudging and discolouration artefact
$I_{mc}(p)$ = Composited image

The cloned region without any discolouration discrepancy is calculated using Eq. (2) [3]:

$$I_c(p) = \begin{cases} I_s(p) + k * r(p) for\ frame\ 0 \\ I_{cf}(p) + k * r(p) for\ frames\ 1,2\ ....N \end{cases} \qquad (2)$$

Where,
$I_c(p)$ = Cloned region without smudging and discolouration artefact
p = Pixel
k = Mean value interpolation coefficient
$I_s(p)$ = Source image
$I_{cf}(p)$ = Contour flow vector
r(p) = Mean value interpolant

Final composite video without discolouration and smudging artefact is calculated using Eq. (3):

$$I_{mc} = \alpha I_c(p) + (1 - \alpha)I_t(p) \qquad (3)$$

Where,
$I_{mc}(p)$ = Final composited image
p = Pixel
$\alpha$ = Opacity channel
$I_c(p)$ = Cloned region without smudging and discolouration artefact
$I_t(p)$ = Target image

The mean value interpolant is calculated for each pixel of the source and target image to eliminate the smudging discrepancies from the blending region. The interpolation at pixel p is calculated using Eq. (4):

$$r(p) = \sum_{i=0}^{m-1} \lambda_i(p) * diff \qquad (4)$$

Where,

r(p) = Mean value interpolant

p = Inner pixel

i = 0, 1, ......, m-1 are points in the boundary area of the cloned image

m = Starting pixel

$\lambda_i(p)$ = Mean value coordinate at pixel p

diff = Colour difference vector

Mean value coordinate at pixel p is calculated using Eq. (5)

$$\lambda_i(p) = \frac{W_i}{\sum_{j=0}^{m-1} W_j} \quad i = 0, 1, m-1 \tag{5}$$

Where,

$\lambda_i(p)$ = Mean value coordinate for pixel p

p = Inner pixel

i = 0, 1, ......, m-1 are points in the boundary area of the cloned image

j = 0, 1, ......, m-1 are points in the boundary area of the cloned image

m = Starting pixel

$W_i$ = Weight of a pixel i

$W_j$ = Weight of a pixel j

The weight of a pixel i is calculated using Eq. (6)

$$W_i = \frac{\tan(\alpha_{i-1}/2) + \tan(\alpha_i/2)}{b_i - p^2} \tag{6}$$

Where,

$\alpha$ = Angle between the boundary $b_i$ and inner pixel $p_i$

$b_i$ = Pixel of boundary area at a point i

p = Inner pixel

**Table 1:** Enhanced multi-layer mean value cloning algorithm  [3]

| Algorithm: Enhanced Multi-Layer mean value cloning Algorithm |
|---|
| Input: pixels of boundary region $\delta_b$, pixels of inside the boundary region $b_{in}$, foreground, background and unknown regions of trimap and the target video frame |
| Output: composite video without any discrepancies. |
| BEGIN |
| Step 1: produce trimap |
| Step 2: calculate the contour flow vector based on trimap propagation method if frame 0 is not equal to frame N |
| Step 3: for every trimap produced in step 2 compute the mean value using Eq. (5). |
| Step 4: cloned image is generated for every video frames from step 2 and 3 |
| Step 5: the discolouration artefact is removed by calculating the mean value interpolant coefficient where k is a value between 0 and 255 |
| Step 6: to remove the smudging artefact in the boundary region, alpha matte calculated using Eq. (3) |
| Step 7: final multi-layered composite video is generated based on the volumetric image synthesisation technique which is represented in Eq. (1) |
| END |

# 3. Proposed Solution

After reviewing a range of techniques for video composition to produce the mixed reality video, the pros and cons of each technique are analysed. Accuracy in terms of overlay error and visualisation error, occlusion, smudging and discolouration artefacts, aliasing distortions, spatial-temporal consistency, processing time, the motion of the object and anatomical structure are the key aspects that need to be considered for real-time remote collaboration.

Among the list of the techniques in the literature review, the enhanced multi-layer mean value cloning algorithm proposed by Venkata et al. [3] is selected as the best solution. This algorithm used the multi-layer volumetric image sysnthesisation method which gives clear visualisation of the hand while doing surgery, as it is very important to the local surgeon to view the expert surgeon hand and follow the guidelines and recovers the occluded area of the target image. Therefore, multi-layer volumetric image synthesisation will increase the accuracy by reducing the overlay error.

In addition, the solution proposed by K. Shakya et at. [4] is considered as the second-best solution. This solution optimized 3D means value coordinate. Mean value cloning using the 3D mean value coordinates can decrease the sawtooth effect due to brightness distortion in the final composite video. Therefore, 3D mean value coordinates will increase the accuracy by reducing visualisation error.

Lastly, our proposed system considered optimized mean value interpolation proposed by J. Wang el at. [22] as the third best solution. This solution added the local smoothing term in the mean value interpolation as an optimization. This term will make the object of interest consistent with the target video frame brightness by smoothing the local inconsistencies between the source and the target video frame. Therefore, the local smoothing term will maintain the spatial consistency which in turn will increase the accuracy of video composition.

The main aim of our work is to produce the mixed reality video by merging the augmented video and virtual image to support the remote collaboration for real time surgery. Our proposed solution, as shown in Figure 2, has two sites: local site and remote site and one final video: the mixed reality video, as explained further in the following part.

***Surgery site:*** On this site, the Computed Tomography (CT) scan data is used. Augmented video is produced through the segmentation of the CT data. Video frames are generated from the videos and sent to the expert site through an internet connection to be merged with the expert surgeon's hand video.

***Expert surgeon site:*** On this site, real-time surgical video is taken using the camera. Then, video frames are generated from this video. As a result, Tri maps are generated from those video frames. A tri map propagation technique is used along with the contour flow method to track the hand motion of the expert surgeon [3]. An adaptive mesh is constructed to the regions selected with CGAL. To speed up this process, the hierarchical boundary sampling method is also implemented which evaluates and improves the pixels of boundary to be processed more in the later stage.

***Mixed reality:*** The optimized 3D mean value coordinates is calculated for every pixel using the modified mean value cloning [4]. The colour difference vector of the source and the target image is computed to remove the discolouration artefacts [3]. Using the mean values and the colour difference vectors, the mean value interpolation is enhanced by adding the local smoothing term which helps to maintain both spatial as well as temporal consistencies between the source and target video frames [22]. The volumetric image synthesisation method is used to recover the area occluded by the hands of the expert surgeon [3]. Finally, composite results are achieved by combining the synthesized image pixels with the target image pixels, alpha mattes are applied to remove the smudging artefact [3].

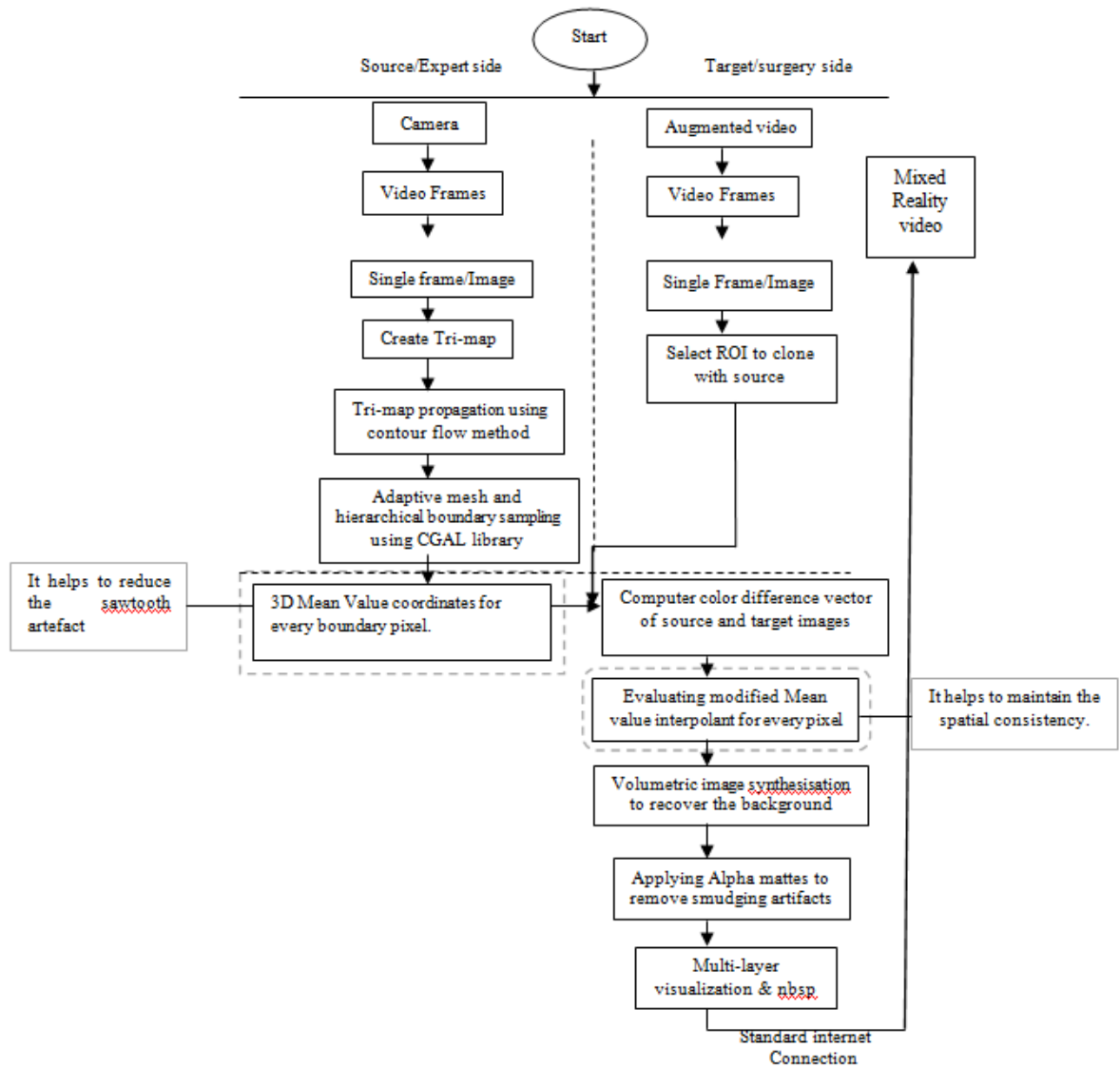

Figure 2: Proposed solution Diagram for Mixed reality visualisation in oral and maxillofacial surgical telepresence

(The green rectangles refer to the modified parts in our proposed system)

$W_i$ is the weight of a pixel I which is represented by the following equation  [4]

$$W_i = \frac{1}{r_i} \sum_{V_i \in (V_i)} \mu_i \qquad (7)$$

Where,

$W_i$ = The weight of a pixel I

i= 0, 1,…… n are the boundary points along the cloned region.

$r_i$ = Space distance between the boundary point and inner point

$$\mu_i = \frac{\beta_{jk} + \beta_{ij} \cdot n_{ij} \cdot n_{jk} + \beta_{ki} \cdot n_{ki} \cdot n_{jk}}{2e_i \cdot n_{jk}} \qquad (7.a)$$

$$n_{jk} = \frac{e_j * e_k}{||e_j * e_k||} \qquad (7.b)$$

Where,

$n_{jk}$ = The unit normal to face (V, $V_j$, $V_k$)

$\beta_{jk}$ = The angle between two segments (V, $V_j$) and (V, $V_k$)

$\beta_{ij}$ = The angle between two segments (V, $V_i$) and (V, $V_j$)

$\beta_{ki}$ = The angle between two segments (V, $V_k$) and (V, $V_i$)

$n_{ij}$ = The unit normal to face (V, $V_i$, $V_j$)

$n_{ki}$ = The unit normal to face (V, $V_k$, $V_i$)

$W_i$ is the weight of a pixel I for the 3D mean value coordinates which is modified from Eq. (6) to the following Eq. [8]:

$$MW_i = \frac{1}{b_i - p^2} \sum_{V_i \in (V_i)} \mu_i \qquad (8)$$

Where,
$MW_i$ = Modified weight of a pixel

$b_i$ = Pixel of boundary area at a point i
p = Inner pixel

When 2D mean value, coordinates are applied in the video composition, unnatural effects may appear in the final merged video due to over-blending and spatial-temporal inconsistencies. To solve these issues, instead of using 2D mean value coordinates, 3D mean value coordinates are calculated [4]. Hence, Eq. (5) is modified to Eq. (9) as follows:

$$M\lambda_i(p) = \frac{MW_i}{\sum_{j=0}^{m-1} MW_j} \ i = 0, 1, m-1 \qquad (9)$$

$M\lambda_i(p)$ = Modified 3D Mean value coordinate for pixel p

$MW_i$ = Modified weight of a pixel

p = Inner pixel
i = 0, 1, ……, m-1 are points in the boundary area of the cloned image
j = 0, 1, ……, m-1 are points in the boundary area of the cloned image
m = Starting pixel

Therefore, the mean value interpolant is modified from Eq. (4) to Eq. (10) as follows:

$$Mr(p) = \sum_{i=0}^{m-1} M\lambda_i(p) * diff \qquad (10)$$

Where,
Mr(p) = Modified mean value interpolant
p = Inner pixel
i = 0, 1, ……, m-1 are points in the boundary area of the cloned image
m = Starting pixel
$M\lambda_i(p)$ = Modified 3D Mean value coordinate for pixel p
diff = Colour difference vector

The cloned region without any discolouration discrepancy is modified from Eq. (2) to Eq. (11) as follows:

$$MI_c'(p) = \begin{cases} I_s(p) + k * r(p) & for\ frame\ 0 \\ I_{cf}(p) + k + Mr(p) & for\ frames\ 1,2, \ldots \ldots N \end{cases} \qquad (11)$$

Where,

$MI_c'(p)$ = Modified cloned region without smudging and discolouration artefact

p = Pixel

$I_s(p)$ = Source image

k = Mean value interpolation coefficient

$I_{cf}(p)$ = Contour flow vector

$Mr(p)$ = Modified mean value interpolant

The mean value interpolant with the local smoothing term is calculated in Eq, (12), as given by J. Wang et at. [22]:

$$Mr'(p) = Mr(p) + k * \sum_{i=0}^{N} H_i * S(i,j) * diff \qquad (12)$$

Where,

$Mr'(p)$ = Modified mean value interpolant with the local smoothing term.

$Mr(p)$ = Modified mean value interpolant

P = Inner pixel

i = 0,1, ..........., m-1 are the boundary points along the cloned region

j = 0,1, ..........., m-1 are the boundary points along the cloned region

diff = Refined colour vector of source and target image

$H_j$ = The normalized weight

S(i, J) = Space distance between two points

k = Parameter controlling the effect of the term

When the colour difference between the source and target frames is bigger, the final composited image seems not natural, which shows the brightness difference between the blending region and the target image. As a result, spatial-temporal inconsistency occurs in the final composited video. Thus, the mean value interpolant is modified by adding local smoothing term from Eq. (10) to Eq. (12) instead of just using Eq. (12), as follows [22]:

$$Mr''(p) = \sum_{i=0}^{m-1} M\lambda_i(p) * diff + k * \sum_{i=0}^{N} H_i * S(i,j) * diff \qquad (13)$$

Where,

$Mr''(p)$ = The final modification of the mean value interpolant.

P = Inner pixel

i = 0,1, ..........., m-1 are the boundary points along the cloned region

j = 0,1, ..........., m-1 are the boundary points along the cloned region

$M\lambda_i(p)$ = Modified mean value coordinate for pixel p

diff = Refined colour vector of source and target image

$H_j$ = The normalized weight

S(i, J) = Space distance between two points

k = Parameter controlling the effect of the term

Besides, the cloned region without any discolouration discrepancy is modified from Eq. (11) to Eq. (14) as follows:

$$MI_c''(p) = \begin{cases} I_s(p) + k * r(p) \, for \, frame \, 0 \\ I_{cf}(p) + k * Mr''(p) \, for \, frames \, 1,2 \, ..... N \end{cases} \quad (14)$$

Where,

$MI_c''(p)$ = The cloned region without smudging and discolouration discrepancy

p = Pixel

$I_s$ = Source image

k = Mean value interpolation coefficient

$I_{cf}(p)$ = Contour flow vector

r (p) = The mean value interpolant.

$Mr''(p)$ = The final modification of the mean value interpolant from Eq. 13.

Furthermore, image composition without smudging and discolouration artefacts and spatial-temporal inconsistencies is modified from Eq. (3) to Eq. (15) as follows:

$$MI_{mc} = \alpha \, MI_c''(p) + (1 - \alpha) \, I_t(p) \quad (15)$$

 Where,

$MI_{mc}$ = The final modified composite image.

P = The inner pixel

$I_t$ = The target image

$\alpha$ = Opacity channel

$MI_c''(p)$ = The cloned region without smudging and discolouration discrepancy

Finally, image multi-layer composition is enhanced from Eq. (1) to Eq. (16) as follows:

$$EI \, layers(p) = \alpha \, MI''_c(p) + \beta \, MI_{mc}(p) \quad (16)$$

 Where,

P = The inner pixel

$MI_{mc}$ = The final modified composite image.

$\alpha, \beta$ = Opacity channel

$MI''_c(p)$ = The cloned region without smudging and discolouration discrepancy.

## 3.1 Area of Improvement

We proposed one equation which is Eq. (16). When the 2D mean value coordinate was applied in the video composition, unnatural effects appeared in the final merged video due to over-blending and spatial-temporal inconsistencies. To solve these issues, instead of using 2D mean value coordinates, 3D mean value coordinates are used and calculated from Eq. (16).  Also, when the colour difference between the source and target frames is bigger, the final composited image seems not natural, which shows the brightness difference between the blending region and the target image, this may lead to spatial-temporal inconsistency in the final composited video. Thus, the mean value interpolant is modified by adding local smoothing as defined in Eq. (10).

Mean value coordinates and image interpolation are the main approaches for video composition. Our proposed system is modifying and enhancing the existing mean value coordinates and image interpolation, as shown in Eq.10, Eq. 12, and Eq. 13. When the colour difference between the object of interest and the target image is very high, the composited video will not seem realistic because the blending object will become inconsistent with the brightness of the target video frames, leading to spatial inconsistencies. Therefore, mean value interpolation is enhanced by adding a local smoothing term to maintain spatial consistency. When 2D mean values coordinates are applied in the video composition, unnatural effects may appear in the final merged video due to over-blending and spatial-temporal inconsistencies. Thus, instead of using 2D mean value coordinates, 3D mean value coordinates are used to overcome the issue of over blending and spatial-temporal inconsistencies. Table 2 shows our proposed algorithm.

Table 2: Proposed algorithm

| Algorithm: Enhanced Multi-Layer mean value cloning Algorithm |
|---|
| Input: pixels of boundary region $\delta_b$, pixels of inside the boundary region $b_{in}$, foreground, background and unknown regions of trimap and the target video frame |
| Output: composite video without any discrepancies |
| BEGIN<br>Step 1: produce trimap<br>Step 2: calculate the contour flow vector based on trimap propagation method if frame 0 is not equal to frame N<br>Step 3: for every trimap produced in step 2 compute the 3D mean value coordinates and enhanced interpolation vector are calculated<br>Step 4: cloned image is generated for every video frame from step 2 and 3<br>Step 5: the discolouration artefact is removed by calculating the mean value interpolant coefficient where k is 0-255<br>Step 6: to remove the smudging artefact in the boundary region, alpha matte calculated using Eq. (15)<br>Step 7: final multi-layered composite video is generated based on the volumetric image synthesisation technique which is represented in Eq. (16)<br>END |

## 4. Results and Discussion

MATLAB R2019b is used to implement the proposed solution. The same software was used to implement the state of the art solution. We compared the results in terms of overlay accuracy, visualization accuracy and processing time, as given in Table 3 to Table 8. Twenty hard tissue samples of oral and maxillofacial surgery and bowel surgery videos are retrieved from online resources based on different age groups. The samples taken have an average video length of 10 minutes. From those videos, frames are extracted using MATLAB. The average number of frames per video samples was 35 frames per video. Twenty video frames are selected from the video frames extracted. The resolutions of the images selected for implementation are 640 by 360 pixels. The samples used for implementation are selected from two different stages of the surgery i.e. intra-operative stage of maxillofacial and bowel surgery and pre-operative stage of the jaw surgery. The reason for selecting these two types is to show the clarity and visualization of the hand on the patient body before and after cutting.

The guidelines from the expertise surgeon can be obtained in the remote areas in the form of virtual hand merged with the augmented video. To understand the guidelines, the virtual hand of expertise surgeon and images of the surgery needs to be visualized clearly. Venkata et al. [3] used the multi-layer mean value cloning algorithm to produce the composite videos but still, the final composite image suffers from spatial-temporal inconsistencies. To overcome this, we implemented 3D mean value coordinates and added the smoothing term in the mean value interpolant. It improves the overlay and visualization accuracy by maintaining the spatial-temporal consistency in the final merged image and minimizes the processing time. Pre-operative mixed view for oral surgery is presented in Figure 3 and intraoperative mixed view for maxillofacial surgery and bowel surgery is presented in Figure 4 and Figure 5 respectively. Figure 3(a), Figure 4(a) and Figure 5(a) are showing the patient images which are obtained from augmented videos. Figure 3(b), Figure 4(b) and Figure 5(b) are showing the hands of the surgeon providing the guidelines. Figure 3(c), Figure 4(c) and Figure 5(c) are showing the mixed reality view to guide the local surgeon throughout surgery for the oral, maxillofacial and bowel surgery respectively.

The samples selected are implemented using MATLAB for both states of the art and proposed solution, and the results obtained from the implementation are shown in Table 3 to Table 8. Each sample is processed and for each of them, accuracy in terms of overlay error, processing time and accuracy in terms of visualization error are computed which are represented in the bar graphs of Figure 6 and Figure 7 respectively. The samples selected have pre-operative as well as intra-operative samples. The difference between the pixel value of the original image and the final merged image gives the overlay errors which are shown in Tables3, 4 and 5 whereas the difference between the RGB values of the original image and the final merged image gives the visualization error. To calculate the accuracy, the image is loaded in the imtool (), a built-in function of MATLAB which in turn displays the image along with the pixel and RGB values information. So, the pixel value and RGB value for that pixel is noted for all the original image samples used and the final merged image for both the state of the art and proposed solution as shown in Table 6, Table 7 and Table 8. The processing time is calculated using run and time (), the built-in function of MATLAB which shows the time to process the frame for both the state of the art and proposed solution for all the samples and the average is calculated. The alpha value of 0.2 was selected for all the samples which make the background image visible in the final composite image. The accuracy in terms of the overlay error and visualization error is enhanced in the proposed solution as compared to the state of art solution as shown in Figure 6 and Figure 8. Similarly, the processing time is slightly decreased in the proposed solution as compared to the state of art solution as shown in Figure 7.

Table 3: Accuracy of overlay error and processing time results for oral samples

| Sample Number | Sample details | AR Video (Patient images) | Expert hand | State of Art System | | | Proposed system | | |
|---|---|---|---|---|---|---|---|---|---|
| | | | | Processed Sample | Overlay accuracy | Processing time/frame in seconds | Processed sample | Overlay accuracy | Processing time/frame in seconds |
| Oral surgery (Intra-operative samples) | | | | | | | | | |
| 1 | Lower front teeth | 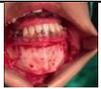 | 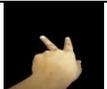 | 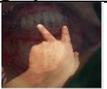 | 1.05mm | 0.23 | 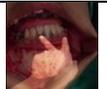 | 0.88mm | 0.156 |
| 2 | Lower teeth | 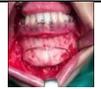 | 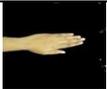 | 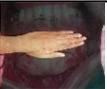 | 0.91mm | 0.224 | 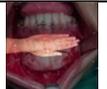 | 0.75mm | 0.163 |
| 3 | Lower front gums | 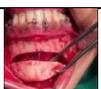 | 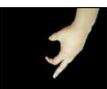 | 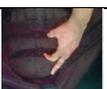 | 1.05mm | 0.221 | 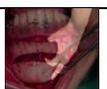 | 0.95mm | 0.175 |
| 4 | Lower gingiva | 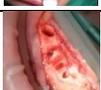 | 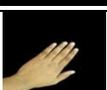 | 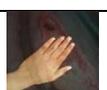 | 1.02mm | 0.22 | 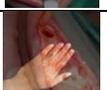 | 0.79 mm | 0.165 |
| 5 | Uppersid e teeth | 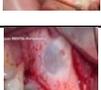 | 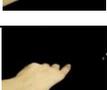 | 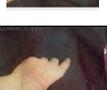 | 0.89mm | 0.227 | 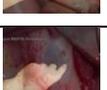 | 0.70mm | 0.188 |
| 6 | Upper gingiva | 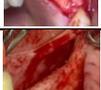 | 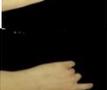 | 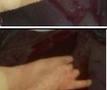 | 1.05mm | 0.229 | 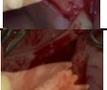 | 0.75mm | 0.205 |
| 7 | Lower gingiva | 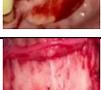 | 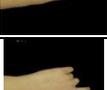 | 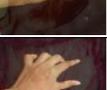 | 1.05mm | 0.22 | 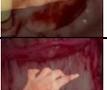 | 0.85mm | 0.185 |
| 8 | Lower gingival | 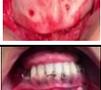 | 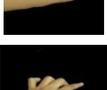 | 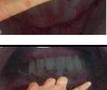 | 1.05mm | 0.21 | 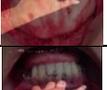 | 0.88mm | 0.190 |
| 9 | Lower molar | 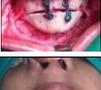 | 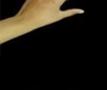 | 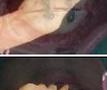 | 1.02mm | 0.2 | 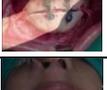 | 0.80mm | 0.177 |
| 10 | Lower gingival | 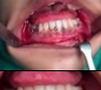 | 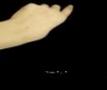 | 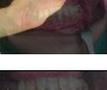 | 0.99mm | 0.21 | 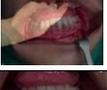 | 0.78mm | 0.159 |
| Average | | | | | 1.008mm | 0.219 | | 0.813mm | 0.176 |

Table 4: Accuracy of overlay error and processing time results for maxillofacial samples

| Sample Number | AR Video (Patient images) | Expert hand | State of Art System | | | Proposed system | | |
|---|---|---|---|---|---|---|---|---|
| | | | Processed Sample | Overlay accuracy | Processing time/frame in seconds | Processed sample | Overlay accuracy | Processing time/frame in seconds |
| Maxillofacial surgery (Pre-operative samples) | | | | | | | | |
| 1 | 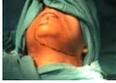 | 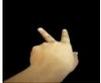 | 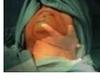 | 1.05mm | 0.214 | 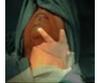 | 0.90mm | 0.156 |
| 2 | 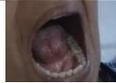 | 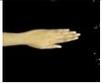 | 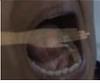 | 0.85mm | 0.225 | 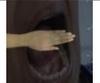 | 0.79mm | 0.163 |
| 3 | 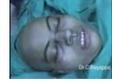 | 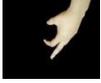 | 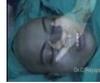 | 1.06mm | 0.207 | 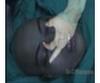 | 0.79mm | 0.175 |
| 4 | 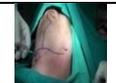 | 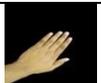 | 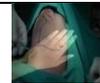 | 1.05mm | 0.246 | 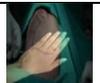 | 0.75mm | 0.165 |
| 5 | 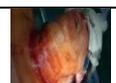 | 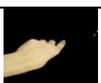 | 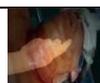 | 0.85mm | 0.198 | 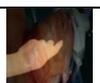 | 0.70mm | 0.188 |
| 6 | 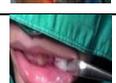 | 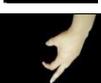 | 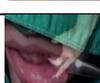 | 0.89mm | 0.2 | 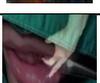 | 0.71mm | 0.205 |
| 7 | 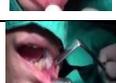 | 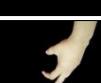 | 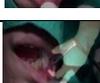 | 1.32mm | 0.204 | 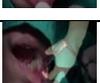 | 0.91mm | 0.185 |
| 8 | 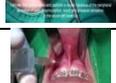 | 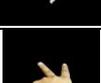 | 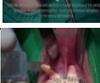 | 1.05mm | 0.21 | 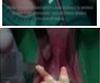 | 0.95mm | 0.190 |
| 9 | 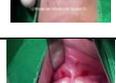 | 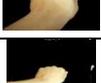 | 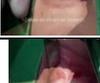 | 1.05mm | 0.195 | 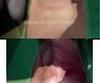 | 0.85mm | 0.177 |
| 10 | 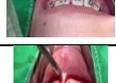 | 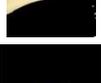 | 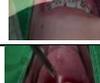 | 0.98mm | 0.189 | 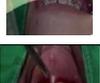 | 0.80mm | 0.159 |
| average | | | | **1.015** | **0.208** | | **0.815** | **0.1763** |

Table 5: Accuracy of overlay error and processing time results for Bowel samples

| Sample Number | AR Video (Patient images) | Expert hand | State of Art System | | | Proposed system | | |
|---|---|---|---|---|---|---|---|---|
| | | | Processed Sample | Overlay accuracy | Processing time/frame in seconds | Processed sample | Overlay accuracy | Processing time/frame in seconds |
| Bowel surgery (Intra-operative samples) | | | | | | | | |
| 1 | 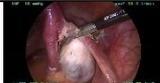 | 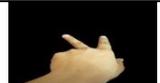 | 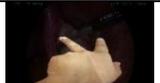 | 0.90mm | 0.21 | 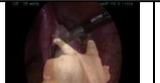 | 0.72mm | 0.150 |
| 2 | 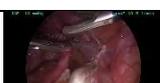 | 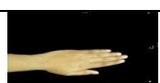 | 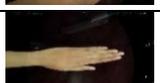 | 1.05mm | 0.220 | 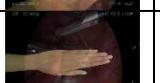 | 0.79mm | 0.165 |
| 3 | 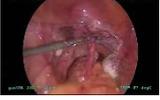 | 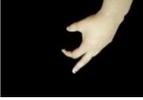 | 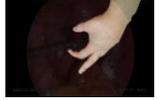 | 1.02mm | 0.205 | 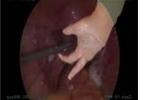 | 0.79mm | 0.185 |
| 4 | 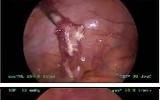 | 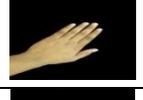 | 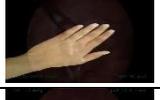 | 1.25mm | 0.214 | 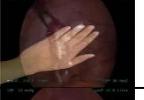 | 0.75mm | 0.168 |
| 5 | 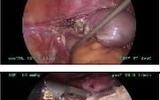 | 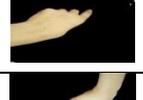 | 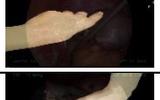 | 1.05mm | 0.2 | 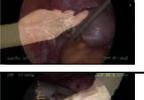 | 0.85mm | 0.175 |
| 6 | 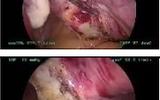 | 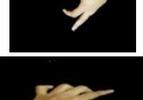 | 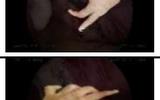 | 0.99mm | 0.2 | 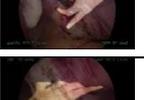 | 0.70mm | 0.163 |
| 7 | 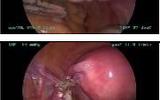 | 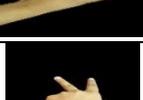 | 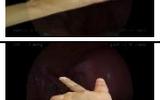 | 1.07mm | 0.21 | 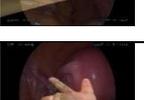 | 0.86mm | 0.171 |
| 8 | 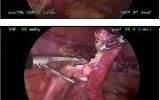 | 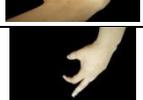 | 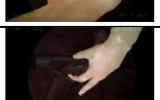 | 0.95mm | 0.195 | 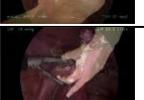 | 0.80mm | 0.155 |
| 9 | 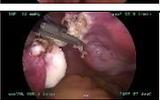 | 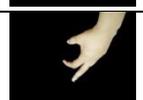 | 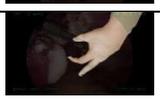 | 0.85mm | 0.215 | 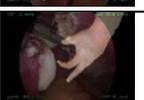 | 0.85mm | 0.177 |
| 10 | 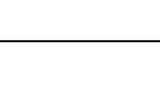 | 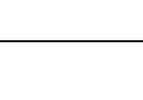 | 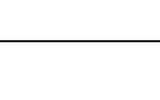 | 1.05mm | 0.202 | 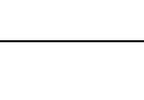 | 0.86mm | 0.180 |
| **average** | | | | **1.018** | **0.207** | | **0.797** | **0.168** |

Table 6: Accuracy of visualization error results for oral samples.

| Sample Number | State of art processed sample | Proposed solution-processed sample | State of Art processed oral sample RGB values | | | The proposed system processed oral sample RGB values | | |
|---|---|---|---|---|---|---|---|---|
| | | | R | G | B | R | G | B |
| 1 | 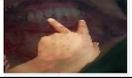 | 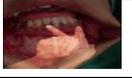 | 222 | 201 | 190 | 200 | 190 | 160 |
| 2 | 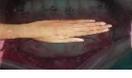 | 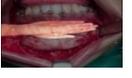 | 236 | 129 | 92 | 170 | 80 | 65 |
| 3 | 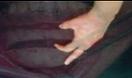 | 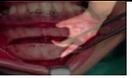 | 211 | 155 | 130 | 180 | 100 | 72 |
| 4 | 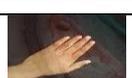 | 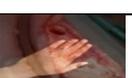 | 225 | 99 | 96 | 205 | 66 | 60 |
| 5 | 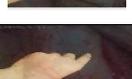 | 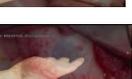 | 231 | 87 | 66 | 190 | 80 | 55 |
| 6 | 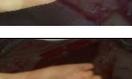 | 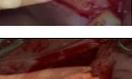 | 236 | 129 | 92 | 197 | 68 | 50 |
| 7 | 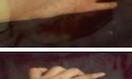 | 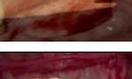 | 227 | 135 | 130 | 187 | 100 | 100 |
| 8 | 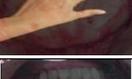 | 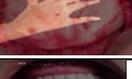 | 245 | 91 | 96 | 177 | 61 | 65 |
| 9 | 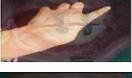 | 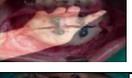 | 203 | 110 | 98 | 160 | 87 | 79 |
| 10 | 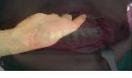 | 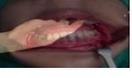 | 210 | 186 | 120 | 180 | 111 | 76 |
| **Average** | | | **224.6** | **132.1** | **111.1** | **184.6** | **94.3** | **78.2** |

Table 7: Accuracy visualization error results for maxillofacial samples.

| Sample Number | State of art processed samples | Proposed solution-processed samples | State of Art processed maxillofacial sample RGB values | | | The proposed system processed maxillofacial sample RGB values | | |
|---|---|---|---|---|---|---|---|---|
| | | | R | G | B | R | G | B |
| 1 | 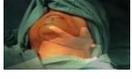 | 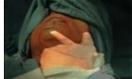 | 145 | 218 | 120 | 110 | 120 | 100 |
| 2 | 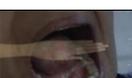 | 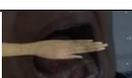 | 120 | 190 | 120 | 105 | 120 | 105 |
| 3 | 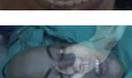 | 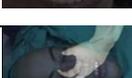 | 150 | 207 | 185 | 130 | 155 | 160 |
| 4 | 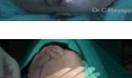 | 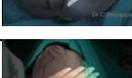 | 188 | 217 | 190 | 150 | 195 | 160 |
| 5 | 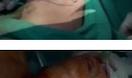 | 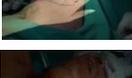 | 220 | 196 | 88 | 190 | 175 | 63 |

| 6 | 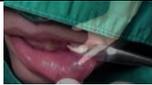 | 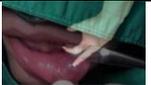 | 208 | 210 | 119 | 185 | 190 | 85 |
|---|---|---|---|---|---|---|---|---|
| 7 | 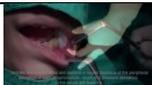 | 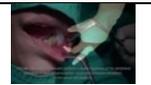 | 210 | 219 | 195 | 188 | 190 | 160 |
| 8 | 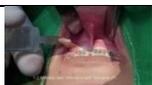 | 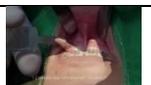 | 217 | 226 | 182 | 160 | 196 | 90 |
| 9 | 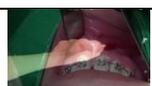 | 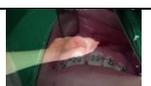 | 230 | 217 | 198 | 201 | 199 | 175 |
| 10 | 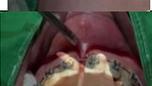 | 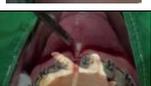 | 225 | 190 | 188 | 200 | 170 | 160 |
| **average** | | | **191.3** | **209** | **158.5** | **161.9** | **171.10** | **125.8** |

Table 8: Accuracy visualization error results for Bowel Samples

| Sample Number | State of art processed sample | Proposed solution-processed sample | State of Art processed oral sample RGB values | | | The proposed system processed oral sample RGB values | | |
|---|---|---|---|---|---|---|---|---|
| | | | **R** | **G** | **B** | **R** | **G** | **B** |
| 1 | 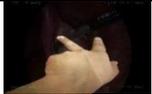 | 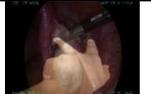 | 220 | 190 | 190 | 190 | 150 | 155 |
| 2 | 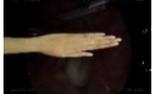 | 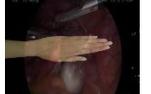 | 225 | 135 | 105 | 165 | 88 | 68 |
| 3 | 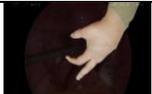 | 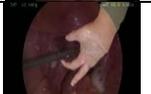 | 222 | 140 | 120 | 178 | 93 | 67 |
| 4 | 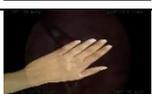 | 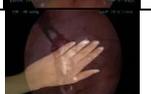 | 217 | 101 | 95 | 179 | 71 | 58 |
| 5 | 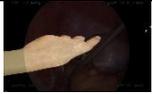 | 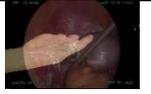 | 226 | 96 | 64 | 190 | 70 | 50 |
| 6 | 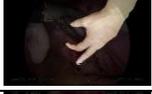 | 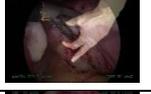 | 230 | 125 | 97 | 180 | 88 | 50 |
| 7 | 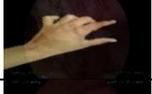 | 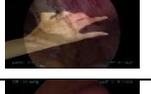 | 218 | 125 | 117 | 175 | 97 | 89 |
| 8 | 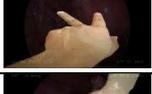 | 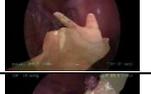 | 220 | 98 | 97 | 177 | 62 | 59 |
| 9 | 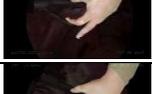 | 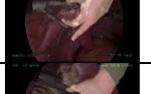 | 205 | 117 | 95 | 160 | 87 | 79 |
| 10 | 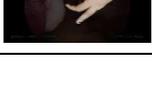 | 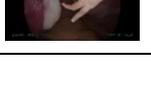 | 211 | 145 | 119 | 175 | 97 | 75 |
| **Average** | | | **219.4** | **127.2** | **109.9** | **176.9** | **90.3** | **70.0** |

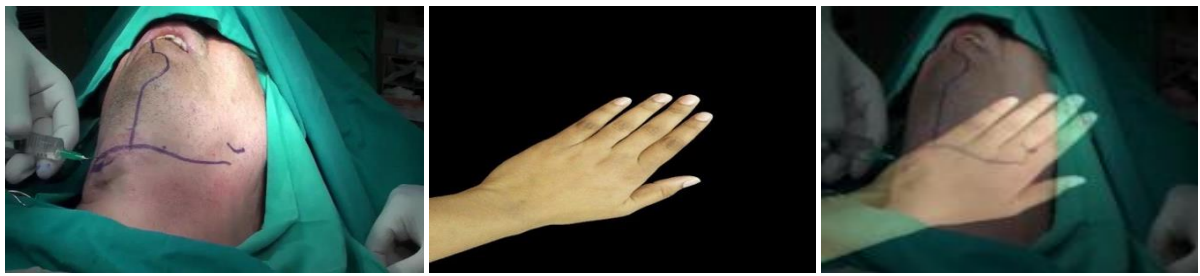

(a)    Augmented image frame            (b)    Expert surgeon hand            (c)    Mixed reality view

**Figure 3**.  The proposed system processed maxillofacial sample

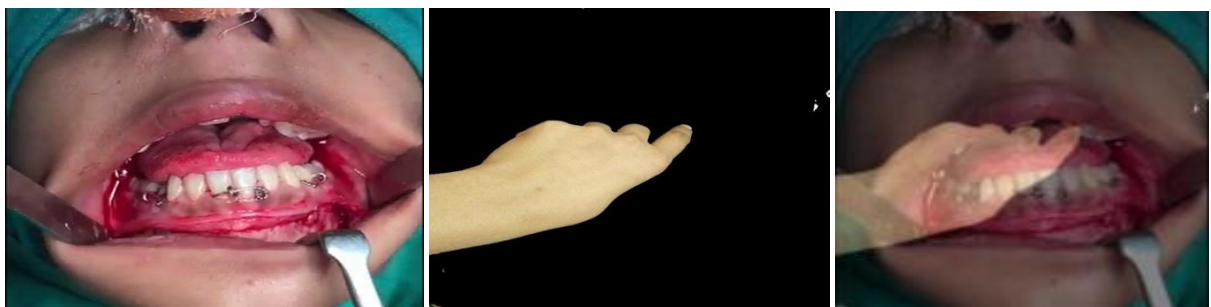

(a)    Augmented image frame            (b)    Expert surgeon hand            (c)    Mixed reality view

**Fig**ure 4 Proposed system processed oral sample.

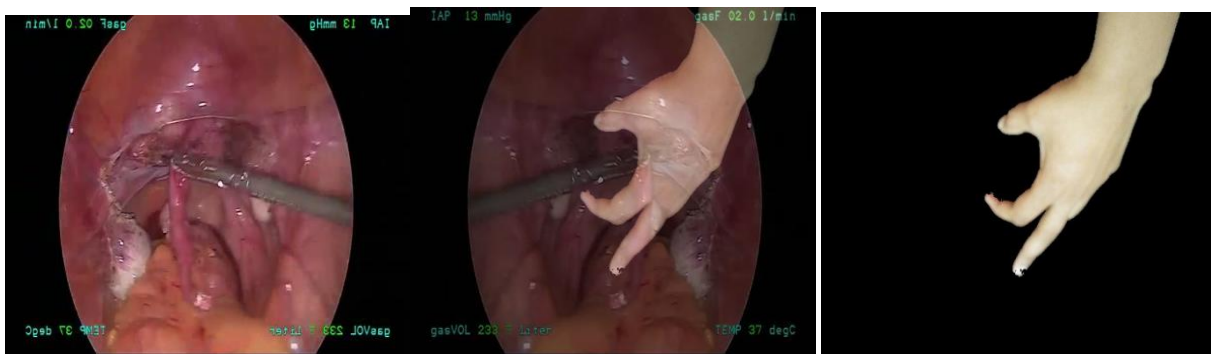

(a)    Augmented image frame            (b)    Expert surgeon hand            (c)    Mixed reality View

**Figure 5** Proposed system processed bowel sample.

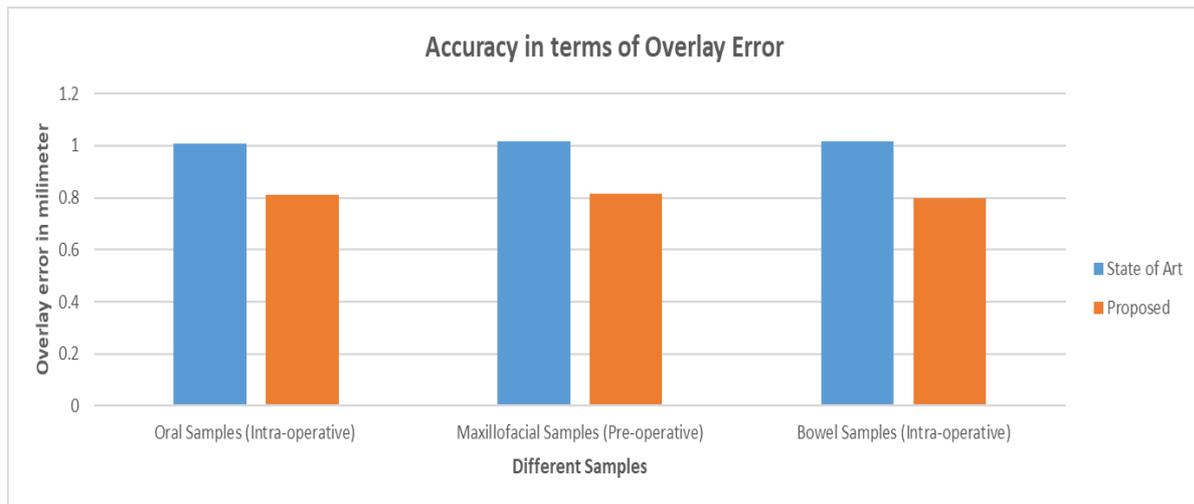

**Figure 6:** Accuracy of overlay error in state of art and proposed solution for oral and maxillofacial samples of pre-operative and intra-operative stages. (a) First two bars showing the overlay error of state of art and proposed solution for oral surgery in the intra-operative stage. (b) The second two bars showing the average overlay error of state of art and proposed solution for maxillofacial surgery in the pre-operative stage. (c) The third two bars showing the average overlay error of state of art and proposed solution for Bowel surgery in the intra-operative stage.

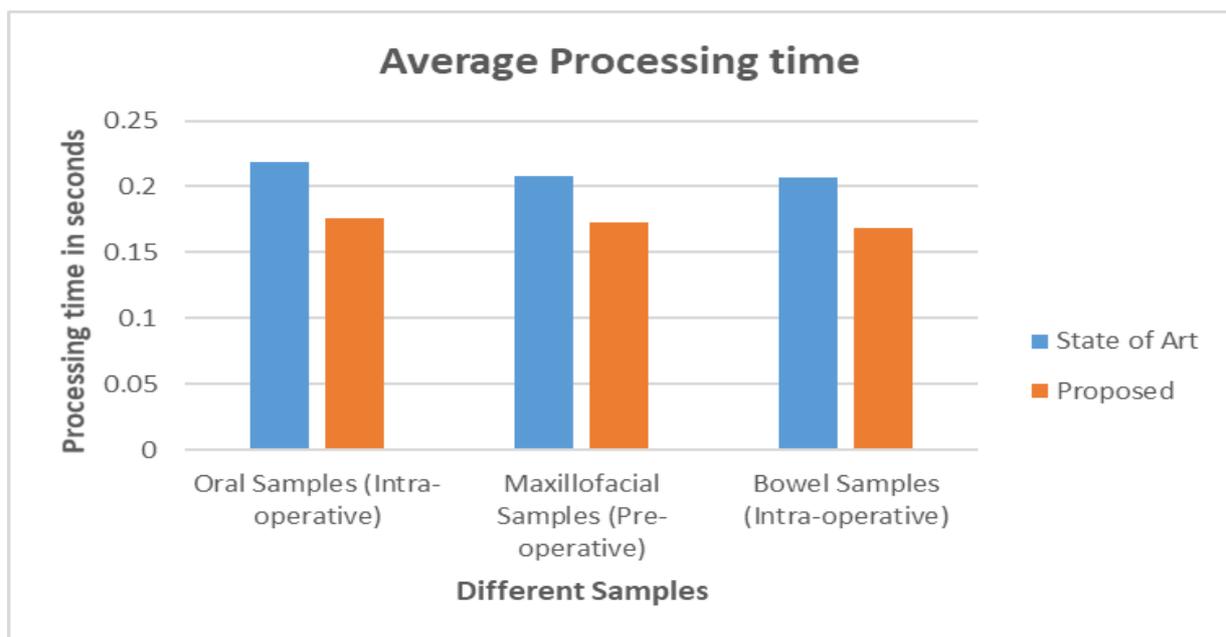

**Figure 7**: Average processing time of all the samples in the state of the art and proposed solution for oral and maxillofacial samples of pre-operative and intra-operative stages. (a) First two bars showing the average of the state of the art and proposed solution for oral surgery in the intra-operative stage. (b) The second two bars showing the average overlay error of state of art and proposed solution for maxillofacial surgery in the pre-operative stage.  (c) The third two bars showing the average overlay error of state of art and proposed solution for Bowel surgery in the intra-operative stage.

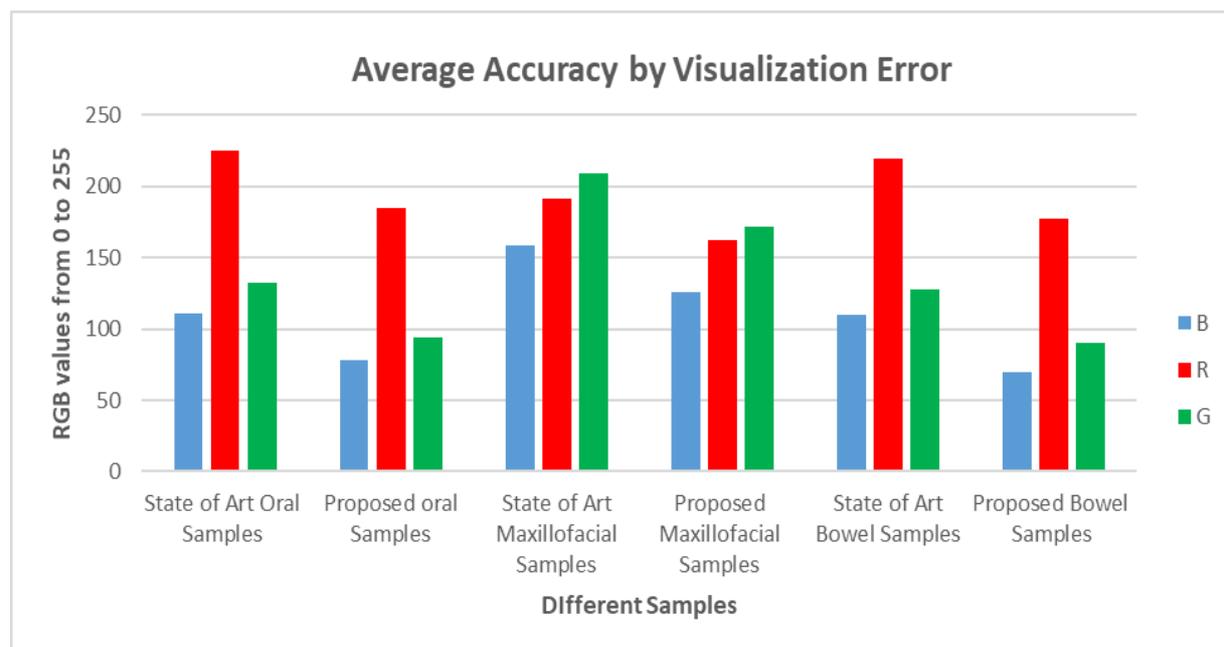

**Figure 8**: Average accuracy by visualization error of all the samples in state of art and proposed solution for oral and maxillofacial samples of pre-operative and intra-operative stages in terms of R (red), G (green) and B (blue) values. (a) First three bars showing the average of RGB values of the state of the art of oral samples. (b) Second three bars showing the average of RGB values of the proposed solution for oral surgery in the intra-operative stage. (c) The third three bars showing the average of RGB values of state of art for maxillofacial sample in the pre-operative stage. (d) The fourth three bars showing the average of RGB values of the proposed solution for maxillofacial surgery in the pre-operative stage. (e) The fifth three bars showing the average of RGB values of state of art solution for bowel surgery in the intra-operative stage. (f) The sixth three bars showing the average of RGB values of the proposed solution for bowel surgery in the intra-operative stage.

The accuracy in terms of overlay error of the proposed solution is improved from 1.01mm to 0.80mm, whereas the accuracy in terms of visualization error is improved from 98.8% to 99.4% compared to our state of the art solution. Similarly, the processing time of the proposed solution is reduced to 0.173 seconds from 0.211 seconds. We have enhanced the mean value cloning algorithm and added the smoothing term to its interpolant, which makes both the background image and surgeon hand visible without the spatial-temporal inconsistencies in the final merged image.

The proposed system can generate the tri map automatically for the rest of the samples once it is generated manually for the first frame using the MATLAB Colour Threshold. The pixel values and RGB values of the image can be calculated using the MATLAB image tool. This function of MATLAB provides the pixel value and RGB value of a point in the image. The difference between the pixel values of the point of the original image and the final composite image gives the overlay error. The pixel values obtained are converted into the millimetre where 1 pixel equals 1.32mm. Similarly, the difference between the RGB values of the point of the image provides the visualization error. The RGB values range from 0-255 where '0' indicates the white colour and '255' indicates the black colour. Besides, the R-value ranges from 200-255 and indicates the red colour, G value ranges from 80-255 and indicates the green colour, and B value ranges from 60-255 and indicates the blue colour. In the proposed system, the colour variations can be adjusted in the final composite image according to the result we want to obtain. The results show that the R-value ranges from 150-210, G value ranges from 70-200 and B value ranges from 50-200. The processing time is automatically calculated by MATLAB once we complete the implementation. In the state of the art solution, there is illumination variation between the background object and the hand of the surgeon in the final composite image.

The tri map propagation is done by using the image tool application of MATLAB. This helps to identify the boundary of the hands accurately. The local smoothing term is added to the mean value interpolant which makes the boundary of the foreground object smooth and the blending results more realistic. The mean value interpolation coefficient helps to reduce the discolouration artefacts. The 3D mean value coordinates are used to remove the illumination variation in the final merged image. These features over-shadowed the limitation of the current solution and enhanced the overlay and visualization accuracy and decreased the processing time. The enhanced 3D mean value cloning algorithm produced the final composite video with an overlay error of 0.80mm, visualization accuracy of 99.4% and processing time of 0.173 seconds.

## 5. Conclusion and Future Work

Mean value coordinates and image interpolation are the main approaches for video composition. Our proposed system is modifying and enhancing the existing mean value coordinates and image interpolation. The proposed solution produced the composite video by merging two videos without illumination variation between the background and foreground object, smudging and discolouration artefact. It overcame the limitations of the current solution by using the 3D mean value coordinates, which helps to minimize the spatial-temporal inconsistencies and adding the local smoothing term. This makes the boundary of the virtual hand smooth and blending result more realistic.

Twenty hard tissue samples of oral and maxillofacial surgery and bowel surgery videos are selected based on different age groups. The samples used for implementation are selected from those videos. Those samples are from two different stages of the surgery: the intra-operative stage of maxillofacial and bowel surgery and the pre-operative stage of the jaw surgery. The reason for selecting these two types is to show the clarity and visualization of the hand on the patient body before and after cutting.

Accuracy is measured in terms of overlay error and visualization error. The difference between the pixel value of the original image and the final merged image gives the overlay errors whereas the difference between the RGB values of the original image and the final merged image gives the visualization error. Our solution minimized the overlay error to 0.80 mm from 1.01 mm and enhanced the visualization accuracy from 98.8% to 99.4%. Processing time is measured as the time taken to implement each frame. The processing time is calculated using the MATLAB function, the time to process the frame for both the state of the art and proposed solution for all the samples and average are calculated. The alpha value of 0.2 was selected for all the samples which make the background image visible in the final composite image. Our solution takes 0.173 seconds to process each frame. The results are obtained by implementing the samples in the MATLAB simulator and compared to the state of the art solution, as shown in Table 9. Since both the surgery scene and the hand of the surgeon giving the guidelines are visible in one frame, it will be easier for the local surgeon to follow the guidelines and perform the smooth surgery. Further study can be done on the automatic generation of the tri map that can make the alpha matting more accurate without working on the individual frames.

Table 9: Comparison Table

| Name of the solution | Proposed Solution | State of Art Solution |
|---|---|---|
| | Enhanced multi-layer 3D mean value cloning algorithm | Multi-layer mean value cloning. |
| Proposed Equation | Final enhanced multi-layer composite image, as shown in Eq. 16 $$EI\ layers(p) = \alpha\ MI''_c(p) + \beta\ MI_{mc}(p)$$ | The multi-layer composition is given as, $$I\ layers\,(p) = \alpha\ I_c(p) + \beta\ I_{mc}(p)$$ |
| Accuracy | Accuracy is measured in terms of overlay error and visualization error. It provides an overlay accuracy of 0.80 mm and visualization accuracy of 99.4% | Accuracy is a measure in terms of overlay error and visualization error. It provides the overlay accuracy of 1.01 mm and visualization accuracy of 98.8% |
| Processing time | Processing time is measured as the time taken to implement each frame. This solution takes 0.173 seconds to process each frame. | Processing time is measured as the time taken to implement each frame. This solution takes 0.211 seconds to process each frame. |
| Contribution 1 | 3D mean value coordinates are used to overcome the issue of over blending and spatial-temporal inconsistencies. | State of art used 2D mean value cloning but it may produce unnatural effects in the final merged video due to over-blending and the spatial-temporal inconsistencies |
| **Contribution 2** | Mean value interpolation is enhanced by adding a local smoothing term to maintain the spatial consistency, as given in Eq. 10, Eq. 12, and Eq. 13 | State of art used mean value interpolation, but it cannot maintain the spatial consistency. |

## Bibliography

[1] M. C. Davis, D. D. Can, J. Pindrik, B. G. Rocque and J. M. Johnston, "Virtual interactive presence in global surgical education: international collaboration through augmented reality," *World neurosurgery,* vol. 86, pp. 103-111, 2016. doi:10.1016/j.wneu.2015.08.0

[2] M. Anvari, "Remote telepresence surgery: the Canadian experience," *Surgical Endoscopy,* vol. 21, no. 4, pp. 537-541, 2007. doi:10.1007/s00464-006-9040-8

[3] H. S. Venkata, A. Alsadoon, P. W. Prasad, O. H. Alsadoon, S. Haddad, A. Deva and J. Hsu, "A Novel Mixed Reality in Breast and Constructive Jaw Surgical Telepresence.," *Computer Methods and Programs in Biomedicine, 177, 253-268. Retrieved from ht,* vol. 177, pp. 253-268, 2019. Retrieved from https://doi.org/10.1016/j.cmpb.2019.05.025

[4] K. Shakya, S. Khanal, A. Alsadoon, P. W. Prasad, J. Hsu, A. Deva and A. . . . Elchouemic, "Remote surgeon hand motion and occlusion removal in mixed reality in breast surgical telepresence: rural and remote care," *American Journal of Applied Science,* vol. 15, no. 11, pp. 497-509, 2019. doi:10.3844/ajassp.2018.497.509

[5] C. Henry and S. W. Lee, "Automatic trimap generation and artefact reduction in alpha matte using unknown region detection," *Expert Systems with Applications,* vol. 133, pp. 242-259, 2019. Retrieved from https://doi.org/10.1016/j.eswa.2019.05.019

[6] H. Jiang, Z. Yuan, M. M. Cheng, Y. Gong, N. Zheng and J. Wang, " Salient object detection: A discriminative regional feature integration approach," *In Proceedings of Computer Vision and Pattern Recognition,* pp. 2083-2090, 2013.

[7] J. Li, G. Yuan and H. Fan, " Generating Trimap for Image Matting Using Color Co-Fusion," *IEEE Access,* vol. 7, pp. 19332-19354, 2019. doi:10.1109/ACCESS.2019.2896084

[8] D. Cho, S. Kim, Y. W. Tai and I. S. Kweon, " Automatic trimap generation and consistent matting for light-field images," *IEEE transactions on pattern analysis and machine intelligence,* vol. 39, no. 8, pp. 1504-1517, 2016.

[9] Z. Q. Cai, L. Ly, H. Huang and Y. H. Liang, "A discrete bio-inspired metaheuristic algorithm for efficient and accurate image matting," *Memetic Computing,* vol. 11, no. 1, pp. 53-64, 2019. Retrieved from https://doi.org/10.1007/s12293-018-0275-4

[10] C. Rhemann, C. Rother and M. Gelautz, " Improving Color Modeling for Alpha Matting," *In BMVC,* vol. 1, no. 2, pp. 1-3, 2008.

[11] B. Amin, M. M. Riaz and A. Ghafoor, " A hybrid defocused region segmentation approach using image matting," *Multidimensional Systems and Signal Processing,* vol. 30, no. 2, pp. 561-569, 2019. doi:https://doi.org/10.1007/s11045-018-0570-8

[12] X. Yi and M. Eramian, "LBP-based segmentation of defocus blur," *IEEE transactions on image processing, 25(4), 1626-1638.,* vol. 25, no. 4, pp. 1626-1638, 2016.

[13] X. Chen, F. He and H. Yu, "A matting method based on full feature coverage," *Multimedia Tools and Applications,* vol. 78, no. 9, pp. 11173-11201, 2019. Retrieved from https://doi.org/10.1007/s11042-018-6690-1

[14] Y. Aksoy, O. A. T. and M. Pollefeys, "Designing effective inter-pixel information flow for natural image matting," *In Proceedings of the IEEE Conference on Computer Vision and Pattern Recognition,* pp. 29-37, 2017.

[15] D. Cho, Y. W. Tai and I. S. Kweon, "Deep convolutional neural network for natural image matting using initial alpha mattes," *IEEE Transactions on Image Processing,* vol. 28, no. 3, pp. 1054-1067, 2019. doi:10.1109/TIP.2018.2872925.

[16] N. Xu, B. Price, S. Cohen and T. Huang, "Deep image matting," *Proceedings of the IEEE Conference on Computer Vision and Pattern Recognition,* pp. 2970-2979.

[17] Z. Fan, J. Lu, C. Wei, H. Huang, X. Cai and X. Chen, " A hierarchical image matting model for blood vessel segmentation in fundus images," *IEEE Transactions on Image Processing,* vol. 28, no. 5, pp. 2367-2377, 2018. doi:10.1109/TIP.2018.2885495

[18] M. M. Fraz, A. Hoppe, B. Uyyanonvara, A. R. Rudnicka, C. G. Owen and S. A. Barman, " An ensemble classification-based appproach to retinal blood vessel segmentation," *IEEE Transactions on BIomedical Engineering,* vol. 59, no. 9, pp. 2538-2548, 2012.

[19] F. J. Lin and J. H. Chuang, "Alpha matting using robust color sampling and fully connected conditional random fields," *Multimedia Tools and Application,* vol. 17, no. 11, pp. 14327-14342, 2018. doi:10.1007/s11042-017-5031-0

[20] Y. Shen, L. Wei, Q. Xu, Z. Peng and Q. Wang, " A simple real-time image cloning algorithm based on modified mean-value coordinates," *International Conference on Control, Automation and Information Sciences,* pp. 118-122, 2015. doi:10.1109/ICCAIS.2015.7338644

[21] Y. Shen, X. Lin, Y. Gao, B. Sheng and Q. Liu, " Video composition by optimized 3D mean-value coordinates," *Computer Animation and Virtual Worlds,* vol. 23, no. 4, pp. 179-190, 2012. doi:10.1002/cav.1465

[22] J. Wang, B. Sheng, P. Li, Y. Jin and D. D. Feng, " Illumination-Guided Video Composition via Gradient Consistency Optimization," *IEEE Transactions on Image Processing,* pp. 1-14, 2019. doi:10.1109/TIP.2019.2916769

[23] T. Chen, J. Y. Zhu, A. Shamir and S. M. Hu, "Motion-aware gradient domain video composition," *IEEE Transactions on Image Processing,* vol. 22, no. 7, pp. 2532-2544, 2013. doi:10.1109/TIP.2013.2251642

[24] S. Pokhrel, A. Alsadoon, P. W. Prasad and M. Paul, "A novel augmented reality (AR) scheme for knee replacement surgery by considering cutting error accuracy," *The International Journal of Medical Robotics and Computer Assisted Surgery,* vol. 15, no. 1, pp. 1-14, 2019. Retrieved from https://doi.org/10.1002/rcs.1958

[25] Y. P. Murugesan, A. Alsadoon, P. Manoranjan and P. W. Prasad, "A novel rotational matrix and translation vector algorithm: geometric accuracy for augmented reality in oral and maxillofacial surgeries," *The International Journal of Medical Robotics and Computer Assisted Surgery,* vol. 14, no. 3, pp. 1-14, 2018. Retrieved from https://doi.org/10.1002/rcs.1889.

[26] B. R. Basnet, A. Alsadoon, C. Withana, A. Deva and M. Paul, "A novel noise filtered and occlusion removal: navigational accuracy in augmented reality-based constructive jaw surgery," *Oral and maxillofacial surgery, 22(4), 385-401. ,* vol. 22, no. 4, pp. 385-401, 2018. Retrieved from https://doi.org/10.1007/s10006-018-0719-5.

[27] J. Wang, H. Suenaga, L. Yang, E. Kobayashi and I. Sakuma, "Video see-through augmented reality for oral and maxillofacial surgery. The International Journal of Medical Robotics and Computer Assisted Surgery, 13(2), 1010-1020. doi:10.1002/rcs.175

[28] Q. Hu, H. Sun, P. Li, R. Shen and B. Sheng, "Illumination-aware live videos background replacement using antialiasing optimization," *Multimedia Tools and Applications,* vol. 77, no. 18, pp. 24477-24497, 2018. Retrieved from https://doi.org/10.1007/s11042-018-5737-7

[29] W. Kim and C. Jung, " Illumination-invariant background subtraction: Comparative review, models, and prospects," *IEEE Access,* vol. 5, pp. 14327-14342, 2019. doi:10.1109/ACESS.2017.2699227

## Appendix

Appendix1- Journal Synopsis

| S.N. | Journal Name | Level of Journal | Journal Format (IEEE or APA) | Link of journals |
|---|---|---|---|---|
| 1. | Computer Methods and Biomedicine | Q1 | IEEE | https://www-sciencedirect-com.ezproxy.csu.edu.au/journal/computer-methods-and-programs-in-biomedicine |
| 2. | American Journal of Applied Sciences | Q2 | APA | https://thescipub.com/journals/ajas |
| 3. | *Expert Systems with Applications* | Q1 | APA | https://www-sciencedirect-com.ezproxy.csu.edu.au/journal/expert-systems-with-applications |
| 4. | *IEEE Access* | Q1 | IEEE | https://ieeexplore-ieee-org.ezproxy.csu.edu.au/document/8289830 |
| 5. | *IEEE Transactions on Image Processing* | Q1 | IEEE | https://ieeexplore-ieee-org.ezproxy.csu.edu.au/xpl/RecentIssue.jsp?punumber=83 |
| 6. | *Memetic Computing* | Q1 | IEEE | https://link-springer-com.ezproxy.csu.edu.au/journal/volumesAndIssues/12293 |
| 7. | *Multidimensional Systems and Signal Processing* | Q1 | APA | https://link-springer-com.ezproxy.csu.edu.au/journal/volumesAndIssues/11045 |
| 8. | *The International Journal of Medical Robotics and Computer Assisted Surgery* | Q2 | IEEE | https://onlinelibrary-wiley-com.ezproxy.csu.edu.au/journal/1478596x?sid=vendor%3Adatabase |
| 9. | *Multimedia Tools and Applications* | Q2 | IEEE | https://link-springer-com.ezproxy.csu.edu.au/journal/volumesAndIssues/11042 |
| 10. | *IEEE Transactions on Image Processing* | Q1 | IEEE | https://ieeexplore-ieee-org.ezproxy.csu.edu.au/xpl/RecentIssue.jsp?punumber=83 |
| 11. | *Oral and maxillofacial surgery* | Q1 | IEEE | https://link-springer-com.ezproxy.csu.edu.au/journal/volumesAndIssues/10006 |
| 12. | *The International Journal of Medical Robotics and Computer Assisted Surgery* | Q2 | IEEE | https://onlinelibrary-wiley-com.ezproxy.csu.edu.au/journal/1478596x?sid=vendor%3Adatabase |
| 13. | *IEEE Transactions on Image Processing* | Q1 | IEEE | https://ieeexplore-ieee-org.ezproxy.csu.edu.au/xpl/RecentIssue.jsp?punumber=83 |

| 14. | *Multimedia Tools and Applications* | Q2 | IEEE | https://link-springer-com.ezproxy.csu.edu.au/journal/volumesAndIssues/11042 |
|-----|--------------------------------------|----|------|------------------------------------------------------------------------------|
| 15. | *Multimedia Tools and Applications* | Q2 | IEEE | https://link-springer-com.ezproxy.csu.edu.au/journal/volumesAndIssues/11042 |